\documentclass[%
reprint,
superscriptaddress,
showpacs,
amsmath,amssymb,
aps,
prx,
longbibliography,
]{revtex4-1}

\usepackage{psfrag,graphicx,epsfig,color}% Include figure files
\usepackage{dcolumn}% Align table columns on decimal point
\usepackage{bm}% bold math
\usepackage{natbib}
\usepackage[usenames,dvipsnames,svgnames,table]{xcolor}
\usepackage{subfigure}
\usepackage{rotating}
\usepackage{float}
\usepackage{lipsum}

\def\re    {{R_\lambda}}
\def\uu {{\mathbf{u}}}
\def\xx {{\mathbf{x}}}
\def\rr {{\mathbf{r}}}
\def\kk {{\mathbf{k}}}

\def\ww {{\boldsymbol{\omega}}}

\def\ff {{\mathbf{f}}}

\def\bi {\begin{itemize}}
\def\ei {\end{itemize}}

\newcommand\ve[1]{\boldsymbol{#1}}

\def\DB#1{{\color{black}{#1}}}

\def\XXint#1#2#3{{\setbox0=\hbox{$#1{#2#3}{\int}$}
     \vcenter{\hbox{$#2#3$}}\kern-.5\wd0}}

\begin{document}

\title{Self-attenuation of extreme events in Navier-Stokes turbulence}

%%% Authors %%%
\author{Dhawal Buaria }
\email[]{dhawal.buaria@ds.mpg.de}
%\thanks{}
\affiliation{Max Planck Institute for Dynamics and Self-Organization, 37077 G\"ottingen, Germany}
\affiliation{Tandon School of Engineering, New York University, New York 11201, USA}
\author{Alain Pumir}
\affiliation{Laboratoire de Physique, Ecole Normale Sup\'erieure de Lyon, Universit\'e de Lyon 1 and Centre National de la Recherche Scientifique, 69007 Lyon, France}
\affiliation{Max Planck Institute for Dynamics and Self-Organization, 37077 G\"ottingen, Germany}

\author{Eberhard Bodenschatz}
\affiliation{Max Planck Institute for Dynamics and Self-Organization, 37077 G\"ottingen, Germany}
\affiliation{Institute for Nonlinear Dynamics, University of G\"ottingen, 37077 G\"ottingen, Germany}

\date{\today}% It is always \today, today,

%\thispagestyle{empty}

\begin{abstract}

Turbulent fluid flows 
are ubiquitous in nature and technology, 
and are mathematically described by the incompressible 
Navier-Stokes equations (INSE).
A hallmark of turbulence is spontaneous generation of
intense whirls, resulting from amplification
of the fluid rotation-rate (vorticity) by its deformation-rate
(strain). This interaction, encoded in the non-linearity of INSE,
is non-local, i.e., depends on the entire state of the flow, 
constituting a serious hindrance in turbulence theory and 
in establishing regularity of INSE.
Here, we unveil a novel aspect of
this interaction, by separating strain into
local and non-local contributions utilizing
the Biot-Savart integral of vorticity 
in a sphere of radius $R$.
Analyzing highly-resolved numerical
turbulent solutions to INSE, we find that when
vorticity becomes very large,
the local strain over small $R$
surprisingly counteracts further amplification.
This uncovered self-attenuation mechanism
is further shown to be connected to local Beltramization
of the flow, 
and could provide a
direction in establishing the regularity of INSE.

\end{abstract}

\maketitle

A parcel of fluid moving at velocity $\mathbf{u} (\xx,t)$ in a flow,
where $\xx\in\mathbb{R}^3$ is the spatial location
and $t$ is time,
simultaneously undergoes rotation and shape deformation,
respectively characterized by the vorticity vector
$\ww = \nabla \times \uu$
and the strain rate
tensor $S_{ij} = (\partial_j u_i + \partial_i u_j)/2$.
Its evolution in time can thereby be described by the
incompressible Navier-Stokes equations (INSE)
written as the vorticity equation \cite{tl72}:
\begin{align}
D_t \omega_i  = \omega_j S_{ij} +  \nu \nabla^2 \omega_i  \ ,
\label{eq:vort}
\end{align}
where $D_t=\partial_t + u_j\partial_j$ is
the material derivative and $\nu$ is the kinematic
viscosity of the fluid.
This equation simply expresses that 
along a parcel trajectory vorticity is non-linearly stretched by the
strain rate, and also subjected to viscous damping.
An essential aspect of this stretching term 
is that it leads to amplification of vorticity,
i.e. generation of enstrophy $\Omega=\omega_i\omega_i$, 
via the production term $P_\Omega = \omega_i\omega_j S_{ij}$,
as readily seen by taking the dot-product 
of Eq.~\eqref{eq:vort} with $\omega_i$ \cite{Tsi2009}.
The rate at which enstrophy is amplified,
and whether it can overcome viscous damping
to blow-up in finite time,
remains one of the outstanding unsolved 
Clay Millennium Prize problems
\cite{Fefferman,doering2009}.

%in experiments and in numerical simulations
%alike~\cite{she_90,zeff:2003,Cardesa,Douady:91},
%which play a pivotal role in numerous natural and engineering
%processes~\cite{falkovich02,durham,ShrSig00}.

It is known that for a finite-time blow-up, $P_\Omega$ must
grow unbounded.
In addition, it has also been proven that this unbounded
growth can possibly only occur when the viscosity $\nu$ 
is sufficiently small \cite{doering2009}, 
which would correspond to turbulent solutions of the INSE.
In fact, it is well known 
that $\Omega$ is highly intermittent in turbulent flows,
attaining values
hundreds or thousands times its mean, becoming 
even more extreme as the  
relative strength of viscosity is decreased 
\cite{Siggia:81, she_90, Douady:91, Jimenez93, Ishihara09, BPBY2019}. 
%Consequently, these extreme events play a 
%pivotal role in many natural and engineering
%processes \cite{collins97,falkovich02,durham,Pitsch2000}.
However, these extreme events are typically found to be 
arranged in tube-like structures
\cite{Siggia:81, she_90, Douady:91, Jimenez93, Ishihara09, BPBY2019},
with geometrical properties deterring 
maximum possible amplification \cite{Ashurst87,jimenez:1992,Tsi2009}.
Nevertheless, 
the question remains open whether the
non-linear amplification could overcome
viscous damping when the flow is sufficiently turbulent.

A fundamental difficulty in analyzing Eq.~\eqref{eq:vort} 
arises from the non-local coupling between vorticity and strain rate;
which implies that strain acting on vorticity at a point,
as in Eq.~\eqref{eq:vort}, is in fact
coupled to the entire state of the flow. 
Specifically, this non-locality can be quantified by expressing the strain tensor
as a Biot-Savart integral of the vorticity field
over the entire 3D spatial domain:
\begin{align}
S_{ij} (\xx) = 
PV  \int_{\xx^\prime} 
\frac{3}{8\pi}  
(\epsilon_{ikl}r_j + \epsilon_{jkl}r_i)  \omega_l(\xx^\prime) \ 
\frac{r_k}{r^5} \ d^3 \xx^\prime \ ,
\label{eq:BS_strain}
\end{align}
where $\rr=\xx - \xx^\prime$ (with $r=|\rr|$) and $\epsilon_{ijk}$ 
is the alternating Levi-Civita symbol.
Thus, the amplification of vorticity can be entirely written
in terms of vorticity itself, but the 
above integral
poses a serious mathematical challenge in understanding the
mechanisms encoded in the non-linearity.
In the current work, by evaluating the above integral numerically, 
we provide evidence that
as vorticity is amplified to large values, 
the strain induced locally will ultimately act 
to attenuate its further amplification.

In order to extract the local strain induced from 
vorticity amplification,
we consider the following decomposition,
by splitting the integration domain
into a spherical neighborhood of radius $R$ and the remaining domain
\cite{ham_pre08,ham_pof08}:
\begin{align}
S_{ij} (\xx) = 
\underbrace{  
\int_{r>R}   \left[ \cdot\cdot\cdot \right]  d^3 \xx'}_{=S^{NL}_{ij}(\xx,R)} \ + \
\underbrace{  
\int_{r \leq R}  \left[ \cdot\cdot\cdot \right] d^3 \xx'}_{=S^L_{ij}(\xx,R)}  \ ,
\label{eq:decomp}
\end{align}
where $\left[ \cdot\cdot\cdot \right] $ 
denotes the integrand in Eq.\eqref{eq:BS_strain}.
The first term $S^{NL}_{ij}$ is
the non-local or background strain
acting on the vorticity to stretch it, whereas
$S^{L}_{ij}$ is the local strain,
induced by the vorticity in its neighborhood in response
to the stretching. 
Thereafter, the production term can also be decomposed as
as $P_\Omega = P_\Omega^{L} + P_\Omega^{NL}$,
where 
$P^{L,NL}_\Omega = \omega_i \omega_j S^{L,NL}_{ij} $.
\DB{For such a decomposition, explicit bounds on
$P_\Omega^{NL}$ can be established in terms of the total
kinetic energy of the flow \cite{Constantin:96}.
Thus, an unbounded growth of $P_\Omega$ 
is only possible through $P_\Omega^{L}$.
However, our results will demonstrate, that 
when $R$ is small enough,
the term $P_\Omega^L$ remarkably acts
to attenuate extreme vorticity fluctuations.
Further analysis reveals that this
attenuation is also connected to
local Beltramisation of the flow, i.e.,
preferrential alignment of vorticity
with velocity, which is expected to
deplete the growth of non-linearity \cite{MoffTsi92}.
}

To analyze the complex interaction between strain and vorticity,
we utilize our unique database generated through 
direct numerical simulations (DNS) of the INSE.
The simulations correspond to 
canonical setup of forced homogeneous and isotropic turbulence
in a periodic domain \cite{Ishihara09}, 
and are performed using the 
well-known Fourier pseudo-spectral methods,
thus allowing us to obtain any quantity of interest
with highest accuracy practicable \cite{mm98}. 
\DB{
It is instructive to note that 
that the mathematical results typically obtained in 
$\mathbb{R}^3$ can be readily generalized to our 
simulation in the $\mathbb{T}^3$ torus}.
%Note that while the numerical results are obtained in $\mathbb{T}^3$,
%they can be  expected to hold in $\mathbb{R}^3$, or alternatiely,
%the mathematical results obtained in $\mathbb{R}^3$, can be generalized
Using the largest grid sizes currently feasible in turbulence
simulations, of up to
$12288^3$ points \cite{Ishihara16,BS2020},  
the Taylor-scale Reynolds number $\re$, which quantifies the
turbulence intensity, is varied from $140$ to $1300$
in our simulations
(corresponding to fully developed turbulence). 
Special attention is given to faithfully resolve the small-scales
and hence the extreme events \cite{BPBY2019},
keeping the grid spacing smaller than 
the Kolmogorov length scale, 
\DB{
$\eta = (\nu^3/\langle \epsilon \rangle)^{1/4}$,
based on the mean dissipation rate of kinetic energy  
$\langle \epsilon \rangle$, 
where the average $\langle \cdot \rangle$ is
taken over the 3D spatial domain and also multiple realizations.
Note that the mean enstrophy 
$\langle \Omega\rangle$, is equal to $\langle \epsilon \rangle/\nu$, 
due to underlying homogeneity \cite{tl72}.}
Additional details about our DNS
and database are provided in the Methods section.

\section*{Results}

\paragraph*{\bf Efficient determination of the local and non-local strain:}
While the vorticity and strain fields can be easily obtained 
from DNS, we have devised an efficient method
to compute the local and non-local strain fields,
without directly evaluating the prohibitively expensive 
Biot-Savart integral
over the entire domain.
As shown in \cite{ham_pre08}, 
using a Taylor-series expansion of vorticity
over a distance $R$, 
the non-local strain 
$\mathbf{S}^{NL}(\xx,R)$ can be expressed in terms of
the total strain as follows:
\begin{align}\nonumber
S_{ij}^{NL} (\xx,R) = \left[ 1 + 
\frac{R^2}{10} \nabla^2  + \frac{R^4}{280} \nabla^2 \nabla^2 + ... \right. \\ 
\left. + \frac{3R^{2n-2}}{(2n-2)!(4n^2-1)} (\nabla^2)^{n-1} + ... \right]  S_{ij}(\xx)  \ .
\label{eq:S_nl_real}
\end{align}
Starting from the above expression and transforming it
to Fourier space (where the 
differential operator $\nabla^2$ reduces to a simple
multiplication by $-k^2$), leads to the relation
\begin{align}
\hat{S}_{ij}^{NL}(\kk,R) = f(k R) \hat{S}_{ij} (\kk) \ , 
\end{align}
where $\hat{(\cdot)}$ 
denotes the Fourier transform, $\kk$ is the wavenumber vector
with $k=|\kk|$ and $f(k R)$ is an infinite
series. In practice, truncating $f(k R)$ to a finite number of terms 
can at best
provide approximate results \cite{ham_pre08}.
However, as derived in the Supplementary, one can show that 
$f(kR)$ converges to the following expression:
\begin{align}
f(kR) = \frac{3\left[ \sin(kR) - kR \cos(kR) \right] }{(kR)^3} \ .
\label{eq:fkr}
\end{align}
This allows us to evaluate the Biot-Savart integral 
in Eq.\ref{eq:BS_strain} by 
applying a simple transfer function to the total strain rate
in Fourier space, and thus to obtain 
$S_{ij}^{L,NL}$ (and $P_\Omega^{L,NL}$) 
very accurately for any value of $R$.
\DB{
Interestingly, it is worth noting that $f(kR)$ in Eq.~\eqref{eq:fkr}
corresponds to the sinc function in 3D,
which also happens to be the Fourier transform
of a box or top-hat filter (of radius $R$), commonly utilized 
in other disciplines, e.g. large-eddy simulation (LES), 
signal processing. 
Thus, evaluating the non-local strain essentially reduces 
to a filtering operation on the total strain.
}

\begin{figure*}
\begin{center}
\subfigure[\ enstrophy, $\Omega=\omega_i\omega_i$]{
\includegraphics[width=0.29\textwidth]{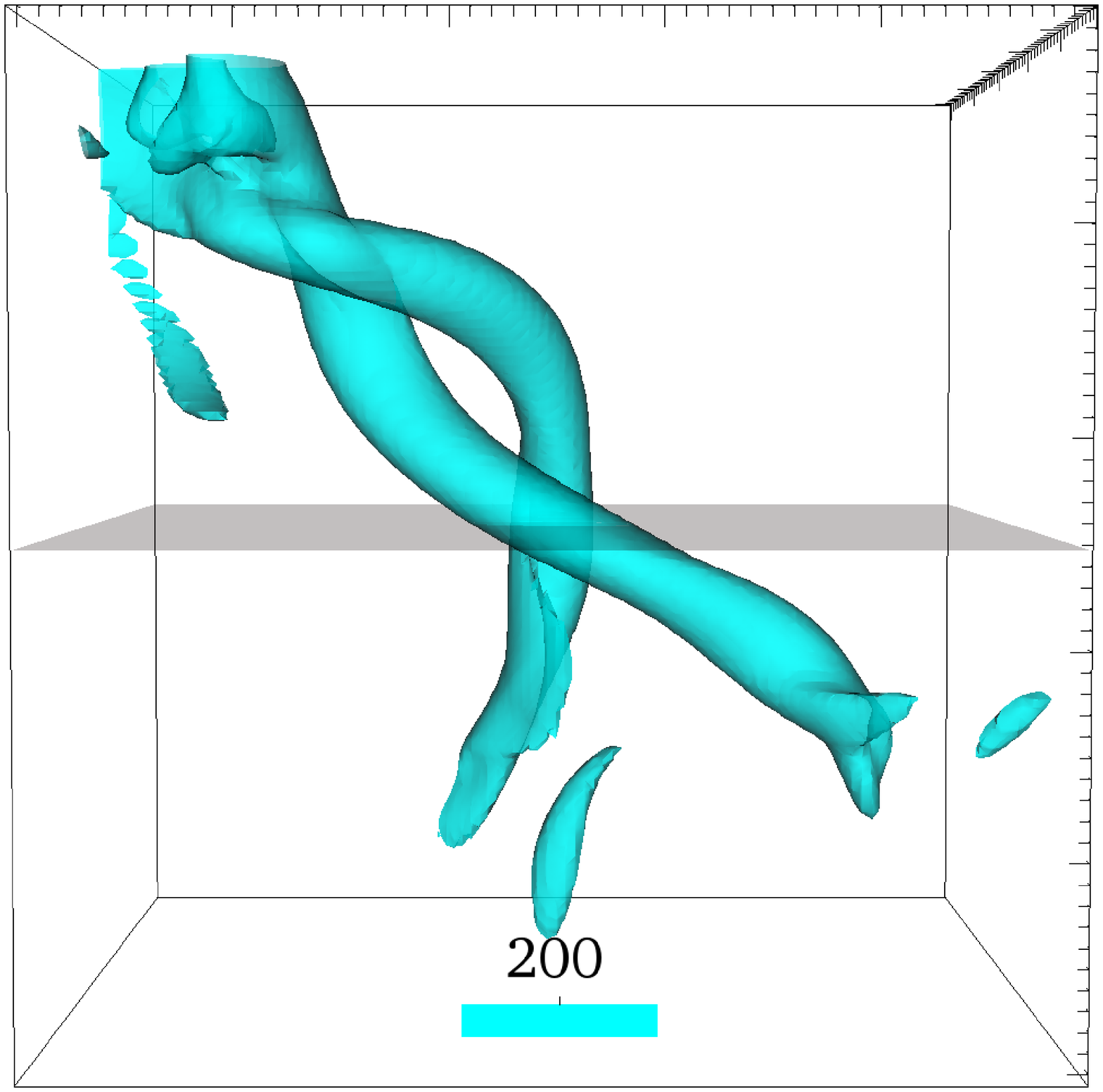}  \ \
}
\subfigure[\ $\Omega$]{
\includegraphics[width=0.29\textwidth]{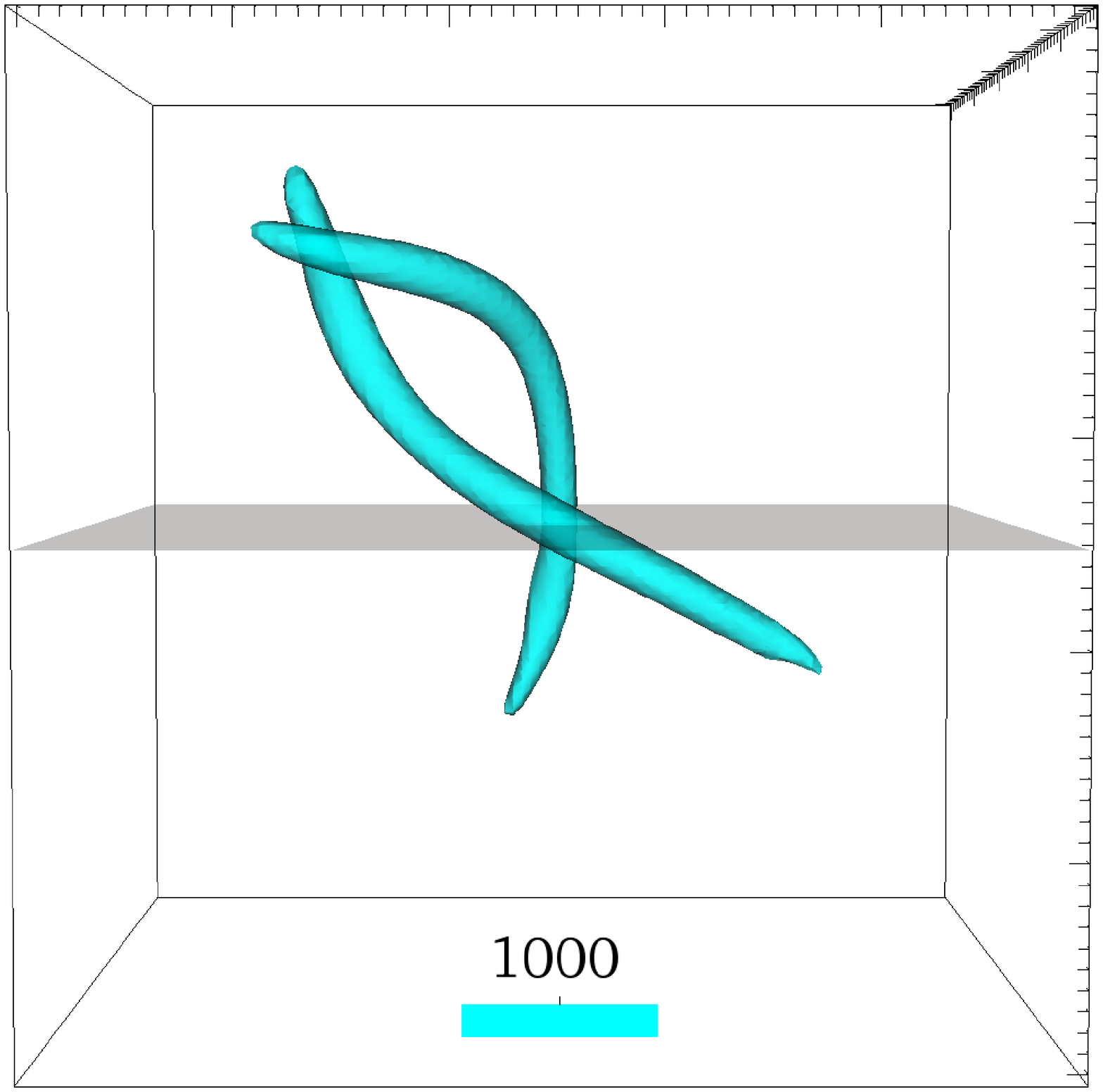} \ \ 
}
\subfigure[\ $\Omega$]{
\includegraphics[width=0.29\textwidth]{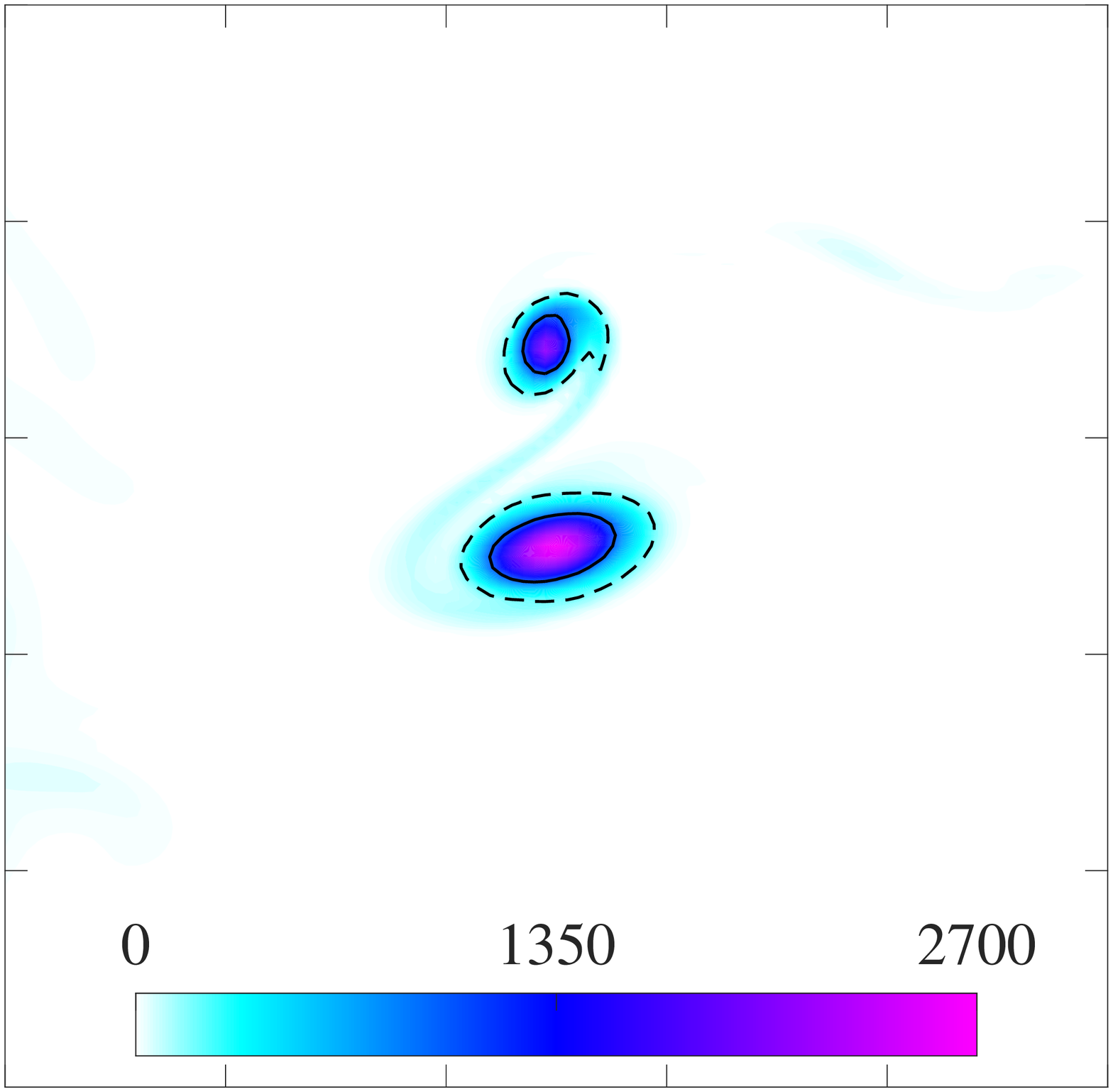} 
} \\
\subfigure[\ enstrophy production based on total strain, $P_\Omega = \omega_i\omega_j S_{ij}$]{
\includegraphics[width=0.29\textwidth]{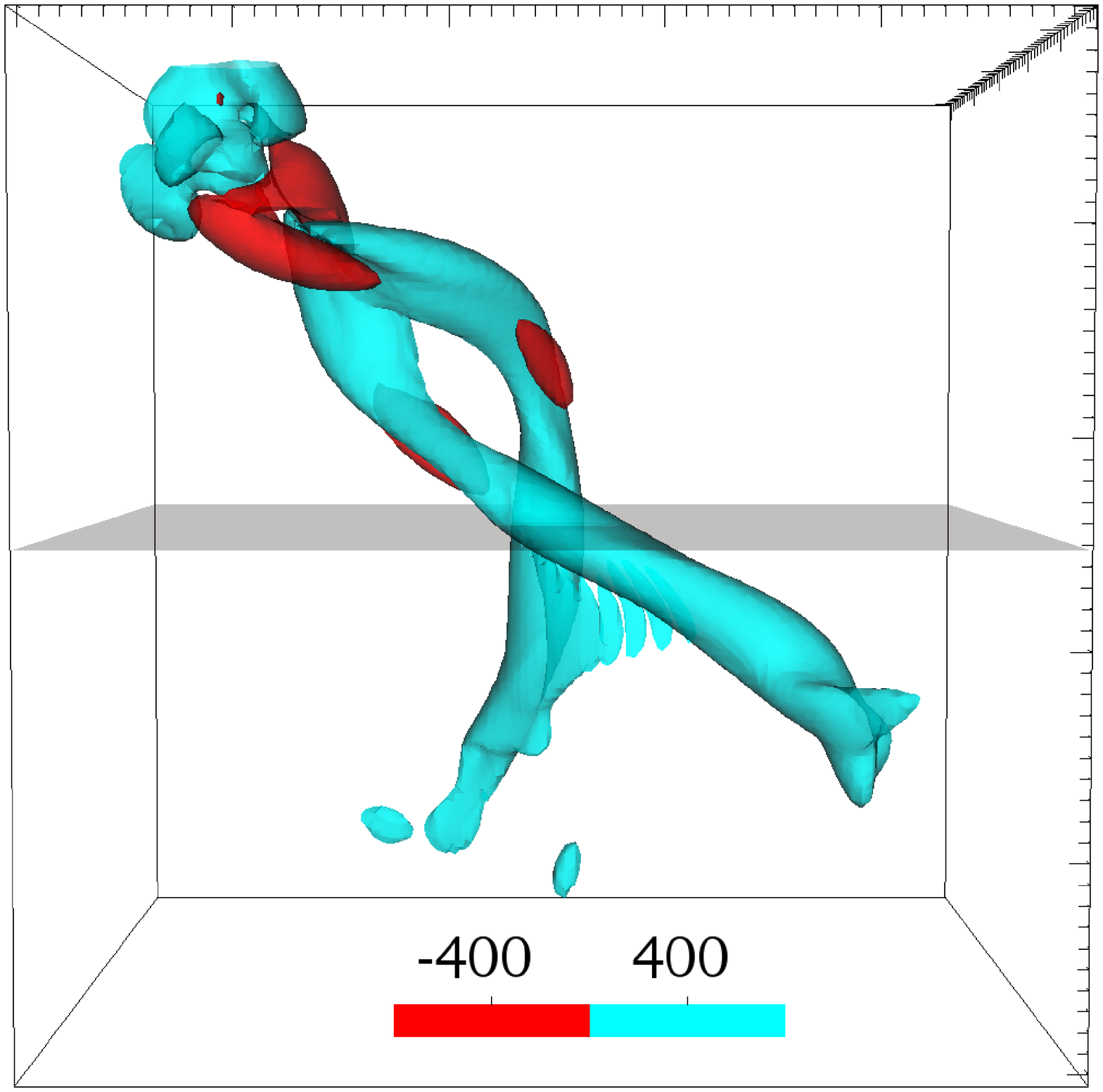}  \ \
}
\subfigure[\ $P_\Omega$]{
\includegraphics[width=0.29\textwidth]{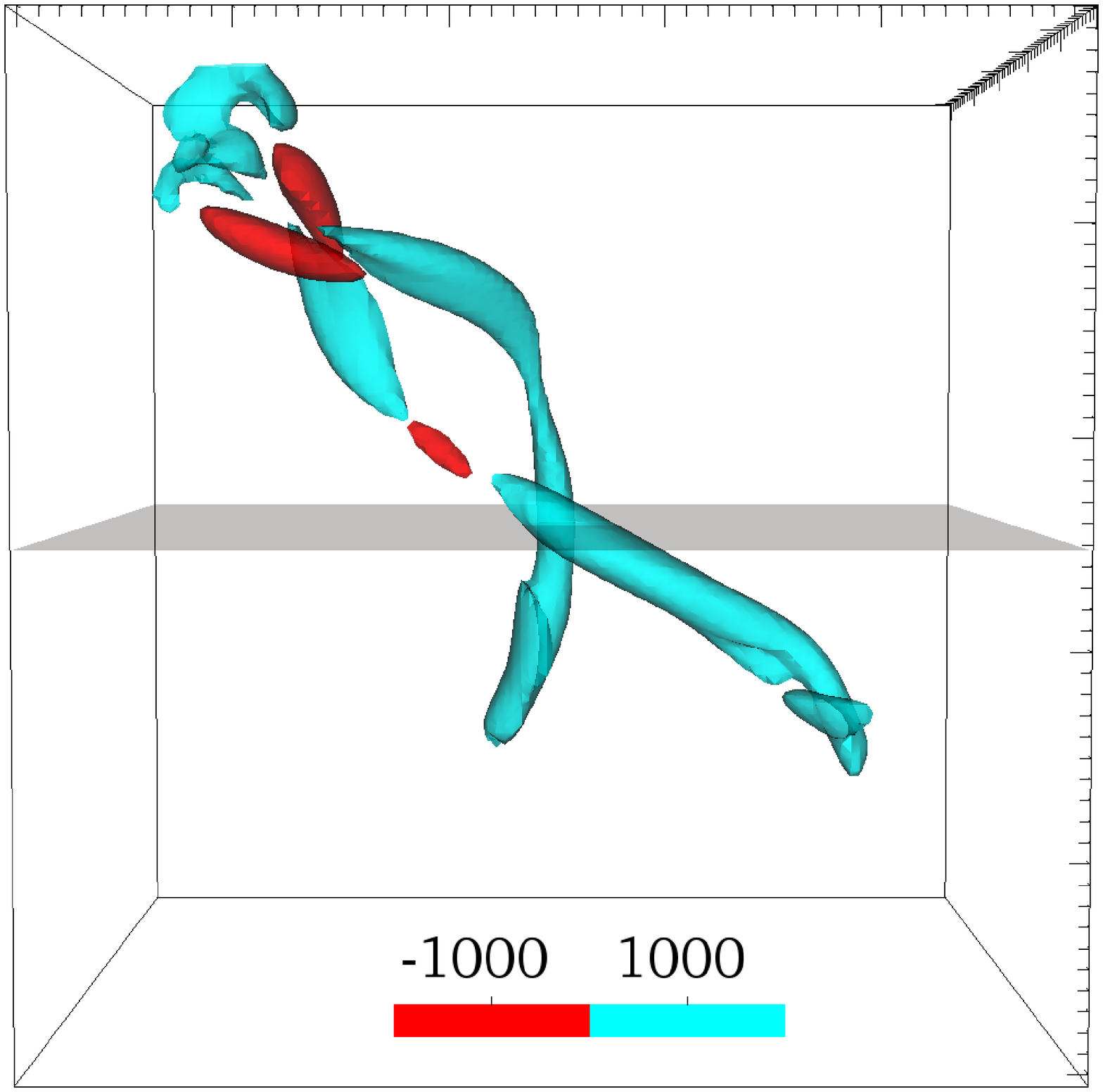}  \ \
}
\subfigure[\ $P_\Omega$]{
\includegraphics[width=0.29\textwidth]{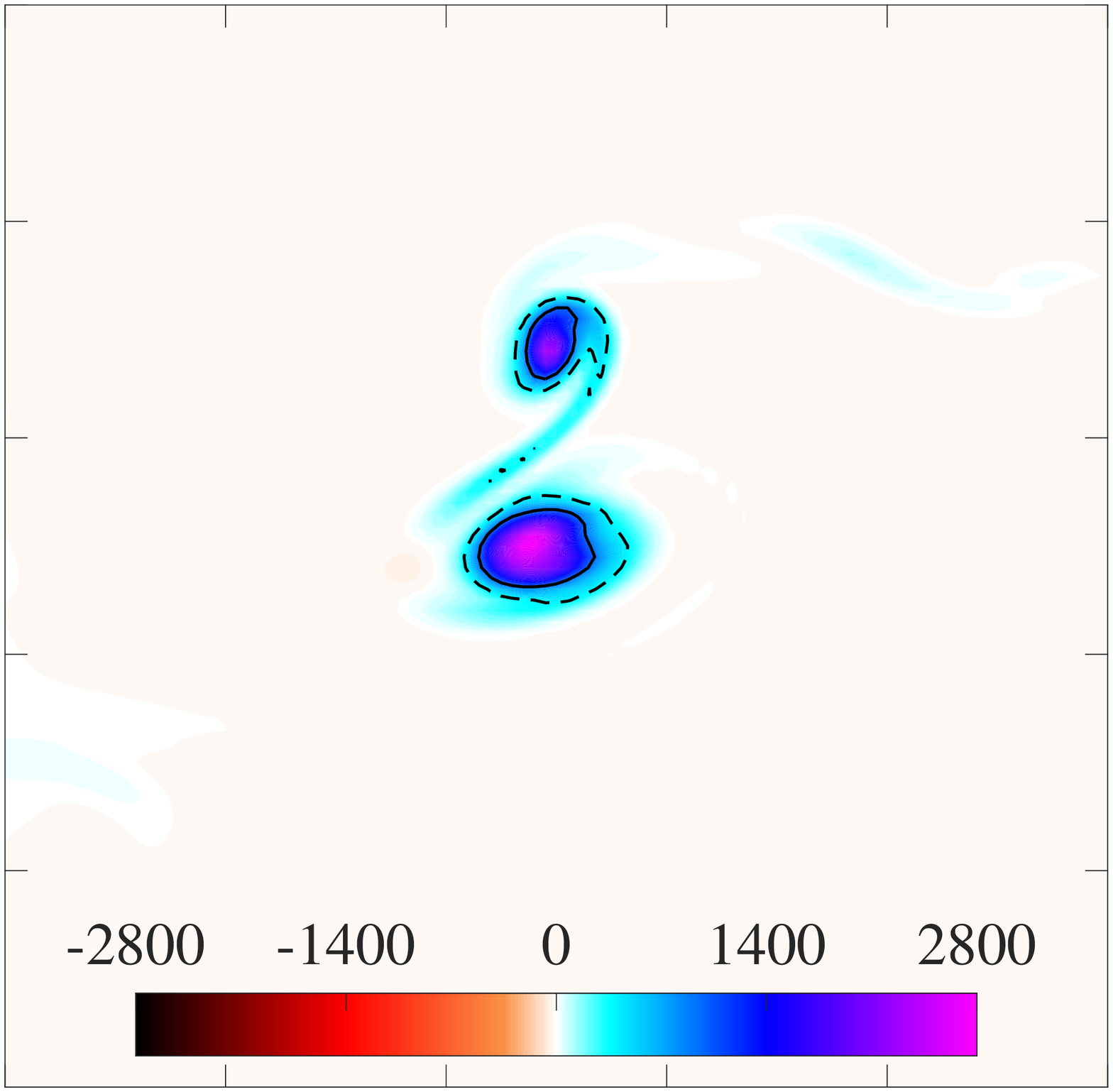} 
} \\
\subfigure[\ enstrophy production based on local strain, $P^L_\Omega = \omega_i\omega_j S^L_{ij}$]{
\includegraphics[width=0.29\textwidth]{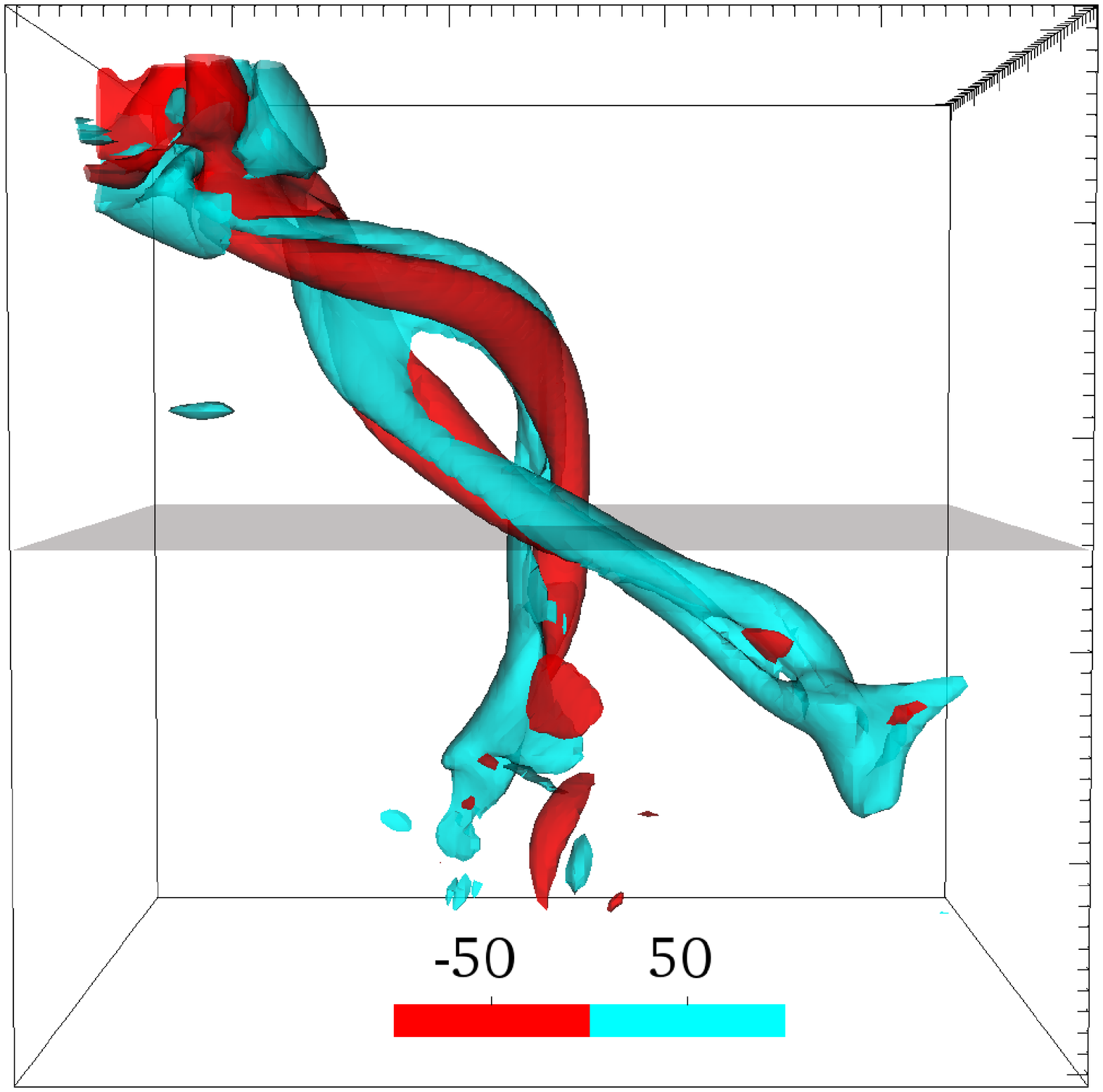} \ \
}
\subfigure[\ $P^L_\Omega$]{
\includegraphics[width=0.29\textwidth]{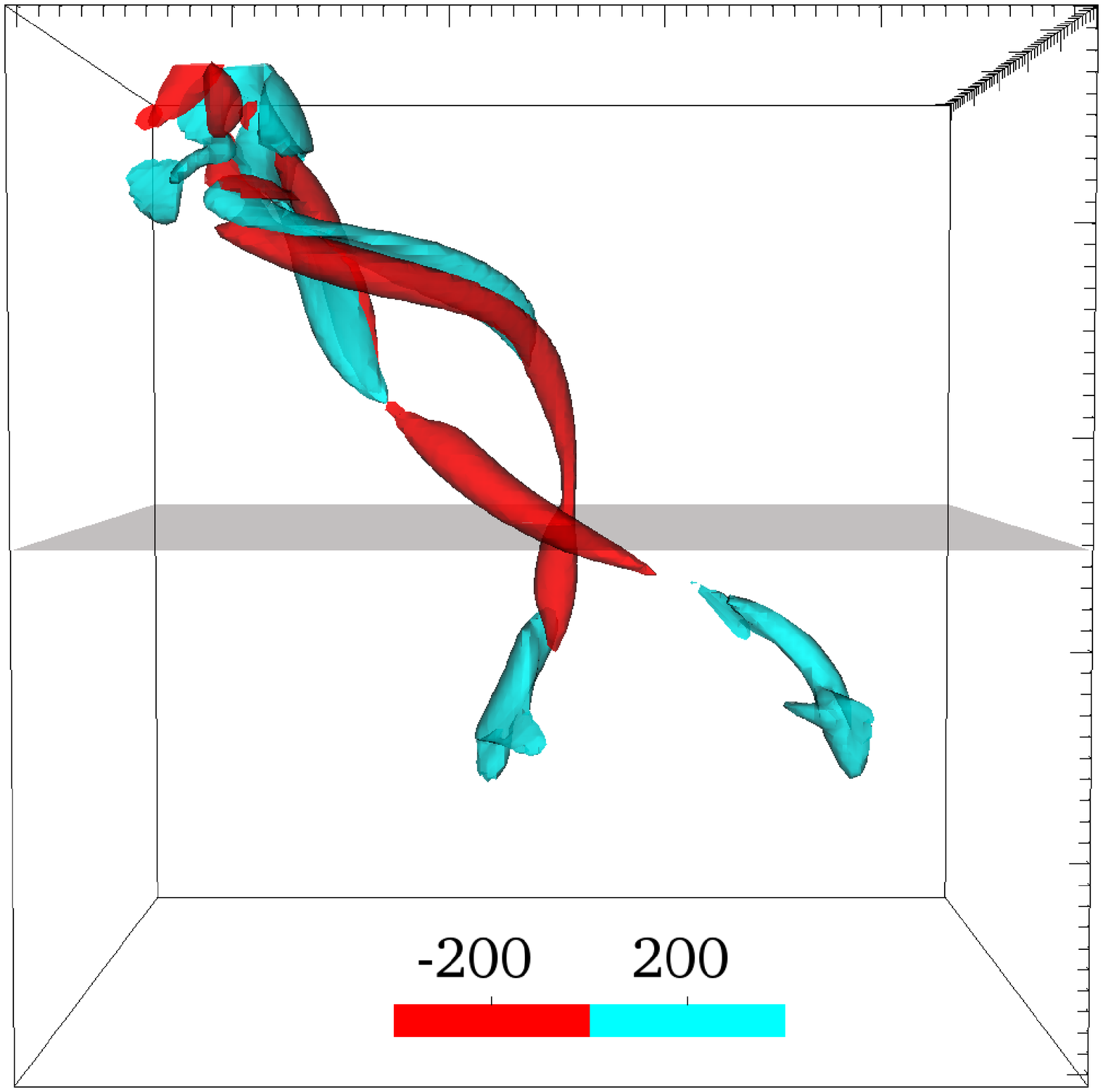} \ \
}
\subfigure[\ $P^L_\Omega$]{
\includegraphics[width=0.29\textwidth]{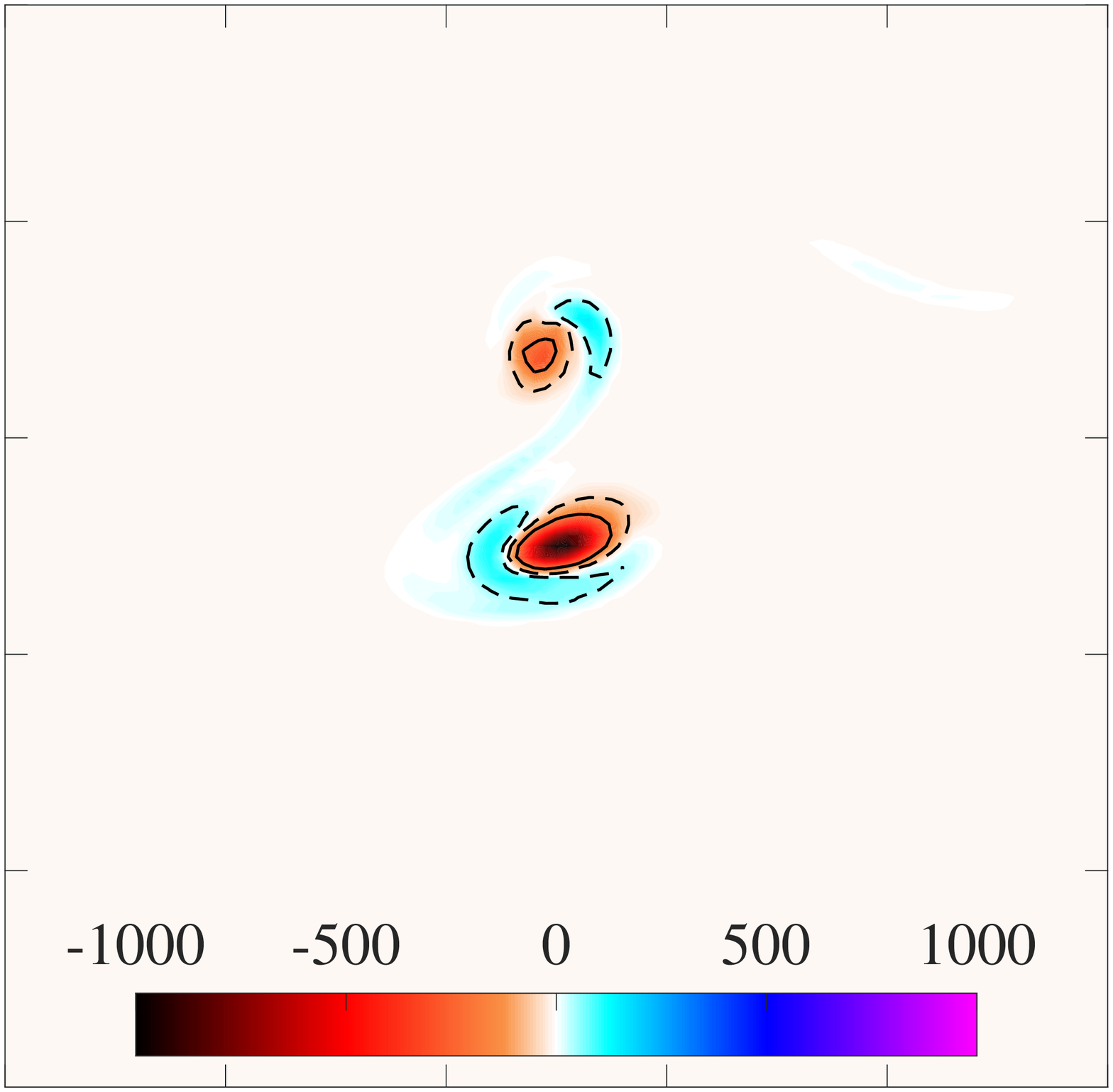} \ \
} \\
\caption{
{\bf Prevalence of negative local stretching in regions of intense vorticity}.
The panels focus on a representative region of intense vorticity
from our numerical simulation at Taylor-scale Reynolds number $\re=650$ 
on a $8192^3$ grid or equivalently of size $(4096\eta)^3$, 
where $\eta$ is the Kolmogorov length scale.
The maximum enstrophy (vorticity-squared) is at the center of the domain
shown, whose edges are $50\eta$ in each direction.
Top row: Isosurfaces of enstrophy at thresholds of 
(a) $200$, and (b) $1000$ (times the mean value).
(c) 2D contours of enstrophy at the mid-plane of the domain, shown in grey in (a) and (b).
Middle row: enstrophy production based on total strain, suitably 
non-dimensionalized by mean of enstrophy, at thresholds of 
(d) $\pm 400$, and (e) $\pm 1000$, which 
approximately correspond to moderate
and intense enstrophy, shown in (a) and (b) respectively.
(f) 2D contours at the mid-plane. The production terms based on total strain
is overwhelmingly positive. 
Bottom row: enstrophy production based on local strain, once again suitably
non-dimensionalized by mean enstrophy, at thresholds of 
(h) $\pm 50$, and (g) $\pm 200$, again corresponding to 
moderate and intense enstrophy  shown in (a) and (b) respectively.
(i) 2D contours at the mid-plane, 
revealing that the
production term based on local strain is strongly negative in the regions of
intense vorticity.
For each row, the thresholds shown in first two 
isosurfaces plots are marked by dashed and solid lines respectively in 
last 2D-contour field plot. 
}
\label{fig:1}
\end{center}
\end{figure*}

\paragraph*{\bf Visualization of extreme events:}
Figure~\ref{fig:1} illustrates our main result,
namely that the 
local contribution to stretching, $P_\Omega^L$,
is in fact {\em negative} in the neighborhood of 
extreme vorticity events.
The visualizations shown in Fig.~\ref{fig:1} focus on a small domain
of size $(50\eta)^3$ around one of the
extreme vorticity events in the flow
(with the most intense vorticity at the center).
Figure~\ref{fig:1}a and b show isosurfaces of 
enstrophy, respectively at 100 and 1000 times the mean value 
corresponding to moderate and intense events,
and illustrate the characteristic vortex-tube structure
\cite{Jimenez93, Ishihara09, BPBY2019}.
The cut through the mid-plane of the domain
is shown in Fig.~\ref{fig:1}c, and
demonstrates the sharp variation 
of enstrophy across the cross section of the tubes.
 
Figure~\ref{fig:1}d-f show the total production $P_\Omega$ for
the same field. 
In Fig.~\ref{fig:1}d, isosurfaces are shown for levels $\pm400$
(with cyan and red corresponding to positive and negative values
respectively), which approximately correspond 
to moderate enstrophy (as shown in Fig.~\ref{fig:1}a).
Whereas in Fig.~\ref{fig:1}e, isosurfaces are shown for $\pm1000$, 
which correspond to intense enstrophy (as shown in 
Fig.~\ref{fig:1}b).
In Fig.~\ref{fig:1}f, the 2D contour field at the mid-plane
is shown. The main observation is that  $P_\Omega$
is overwhelmingly positive, which is anticipated
given large enstrophy in these tubes, 
and also from dynamical constraints of
turbulence \cite{Betchov56}.

Finally, Fig.~\ref{fig:1}g-i shows the contribution $P_\Omega^L$ from
local strain for $R = 2\eta$. 
In Fig.~\ref{fig:1}g and h, isosurfaces are shown for levels
$\pm50$ and $\pm200$ respectively, once again corresponding
to moderate and intense enstrophy events respectively.
Unlike $P_\Omega$ which is always positive on average \cite{Betchov56}, 
the mean of $P_\Omega^L$ has no such constraints. 
For moderate values shown in 
Fig.~\ref{fig:1}g, we find that 
the volumes occupied by positive and negative values
are comparable. However, for intense value shown in
Fig.~\ref{fig:1}h, {\em negative} stretching rate is 
more prevalent
especially around the center where vorticity is maximum.
This is corroborated by
Fig.~\ref{fig:1}i, which shows the 2D contour level of $P_\Omega^L$ at 
the mid plane and reveals that
both negative
and positive values occur
%It is clearly seen that 
in the outer regions of the tubes
where vorticity is not very  intense;
whereas large negative values occur inside the tubes,
where vorticity is most intense. 

\DB{
Let us briefly mention that the flow structure presented in 
Fig.~\ref{fig:1} represents one generic scenario
of how the regions of intense vorticity look like. 
Needless to say, we inspected many such regions, 
and note that all of them qualitatively behave in the same
manner, and essentially lead to the same conclusion.
We have included another such example in the Supplementary.}

\begin{figure}
\begin{center}
\includegraphics[width=0.45\textwidth]{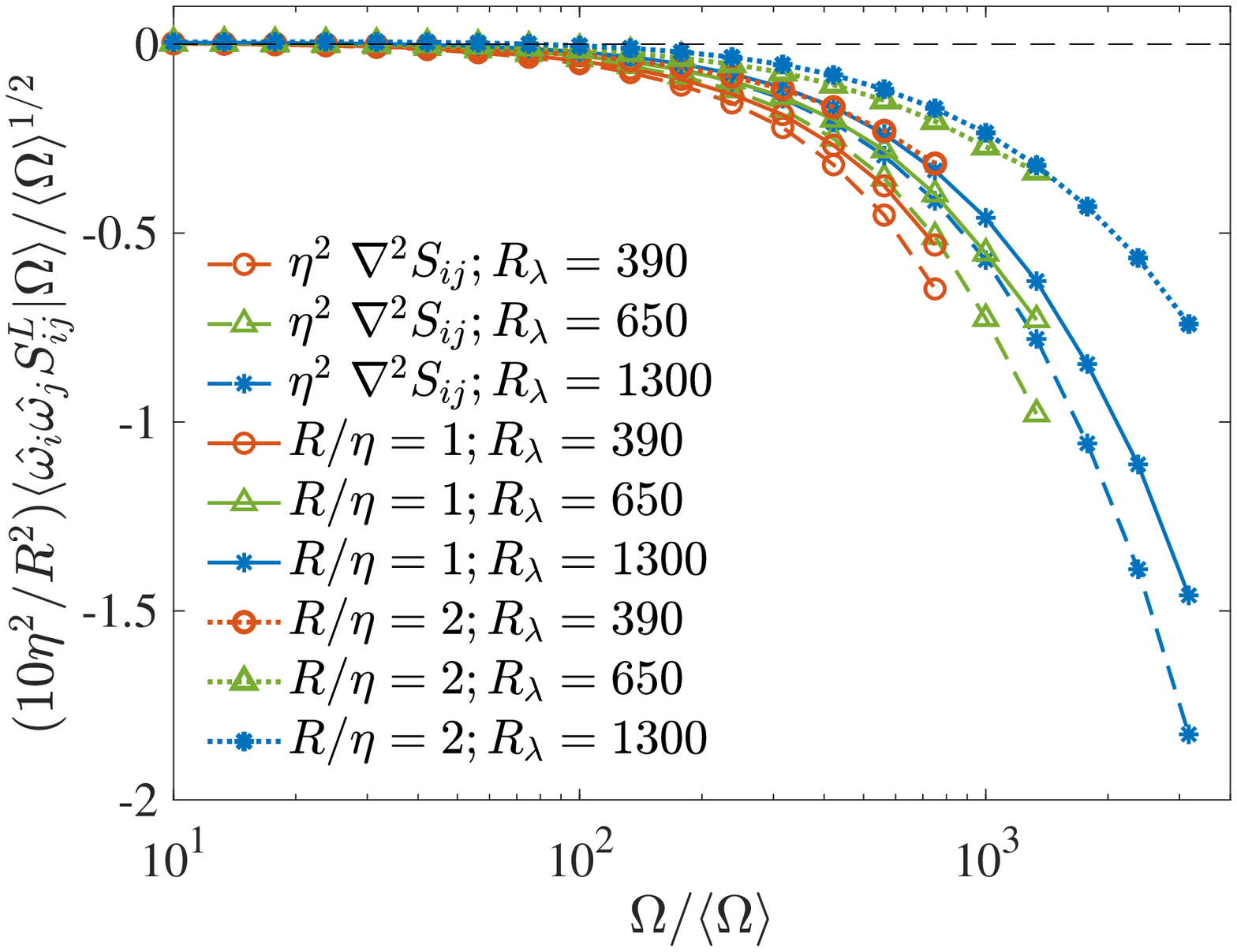} \\ 
\ \ \
\includegraphics[width=0.44\textwidth]{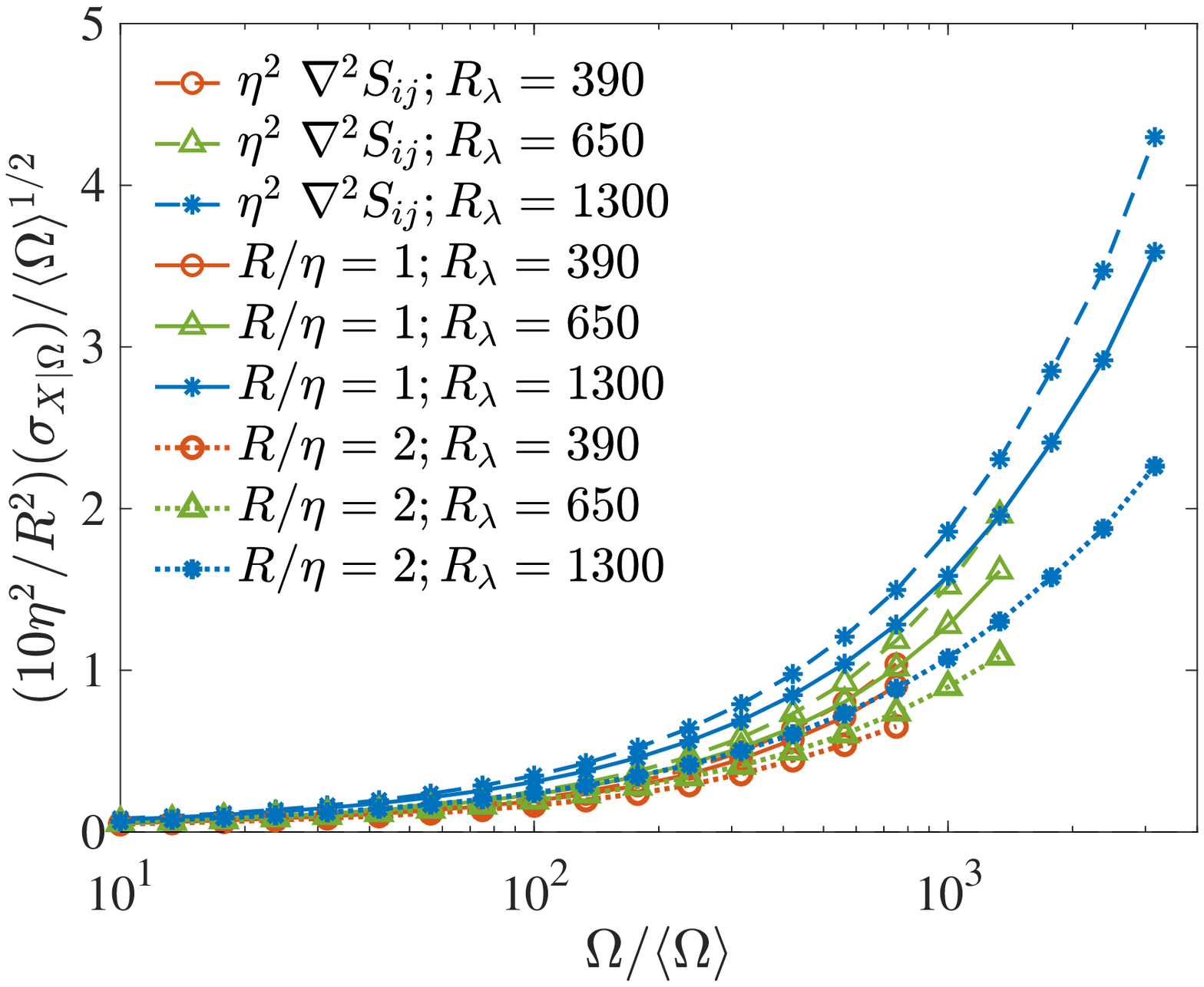}
\caption{
{\bf
Negative contribution of local strain to production of enstrophy}.
(a) Averaged enstrophy production due to the local strain, $P^L_\Omega$,
conditioned on enstrophy normalized by
its mean value.
The curves shown correspond for $R/\eta=1$ and $2$,
at Taylor-scale Reynolds numbers $\re=390-1300$.
For comparison, we also show the contribution based on
$\eta^2 \nabla^2 S_{ij}$, which is the limiting
value of local strain for small $R$ as noted in Eq.~\eqref{eq:slocal}.
(Accordingly the curves for $R/\eta=1$ and $2$ are also
adjusted by a factor of $10\eta^2/R^2$).
\DB{(b) The conditional root-mean-square $\sigma_{X|\Omega}$
of the local enstrophy production term 
($X = \hat{\omega}_i \hat{\omega}_j S^L_{ij}$), 
defined as
$\sigma^2_{X|\Omega} = \langle X^2|\Omega\rangle - \langle X|\Omega\rangle^2$.
Similar normalization as panel (a) is
used.}
}
\label{fig:local}
\end{center}
\end{figure}

\paragraph*{\bf Conditional statistics:}
To establish the quantitative significance
of the observations in Fig.~\ref{fig:1},
Fig.~\ref{fig:local}a shows the average of
$P^{L}_\Omega /\Omega$ conditioned on $\Omega$,
for $R=\eta$ and $2\eta$, and various Reynolds numbers.
Note $P^L_\Omega/\Omega = \hat{\omega_i} \hat{\omega_j} S^L_{ij}$
(where $\hat{(\cdot)}$ is the corresponding unit vector)
and provides the measure of effective strain  engendering
enstrophy production, irrespective of the strength of vorticity
\cite{Ashurst87,Tsi2009}.
The Taylor expansion in Eq.~\eqref{eq:S_nl_real} implies
that for small $R$, $S^L_{ij}$ can be written as
\begin{align}
S^L_{ij}(\xx,R) =  - \frac{R^2}{10} \nabla^2 S_{ij}(\xx) + \mathcal{O}(R^4) \ , 
\label{eq:slocal}
\end{align}
which suggests that the local strain 
is in fact proportional to the Laplacian of the total strain.
Hence for comparison, we have also shown the conditional
expectation 
$\langle \hat{\omega}_i \hat{\omega}_j \eta^2 \nabla^2 S_{ij}|\Omega \rangle$ in 
Fig.~\ref{fig:local}a, and 
$P^L_\Omega$ is accordingly multiplied by $10 \eta^2 /R^2$. 
The conditional production term is virtually zero for small 
to moderate values of $\Omega$ -- consistent with 
strong cancellation 
between negative and positive values seen in Fig.~\ref{fig:1}g. 
However, as $\Omega$ gets larger, 
the expectation 
$\langle P^L_\Omega | \Omega \rangle$ becomes {\em negative}
for all Reynolds numbers
and strongly increases in magnitude with $\Omega$.
We note that the values of $P^L_\Omega$ are
overwhelmingly negative for large $\Omega$, 
as corroborated by the observation (not shown in figure)
that conditional expectations of $|P^L_\Omega|$ and $|-P^L_\Omega|$ 
are virtually equal.
%{In fact, further tests demonstrate that 
%Eq.~\eqref{eq:slocal} provides a very good description of $S^L$, up to ? (?, ?)
%in the sense of the $L^2$ norm for $R = 2 \eta$ ($R = \eta$, $R = \eta/2$).}

\DB{
In addition, in Fig.~\ref{fig:local}b, we have show the
conditional root-mean-square (rms) of the fluctuations of the
$P^L_\Omega$, normalized in the same manner as Fig.~\ref{fig:local}a.
Once again, we have included the corresponding 
curve for $\eta^2 \nabla^2 S_{ij}$ for comparison.
Remarkably, we observed the exact behavior as seen in panel
Fig.~\ref{fig:local}a
(except the curves are all on the positive side, because
the rms is always positive by definition).
At the same time, we note that the
curves in both Fig.~\ref{fig:local}a and b, have 
comparable values, i.e., the mean
and rms are comparable (especially for large $\Omega$). 
This reaffirms that $P^L_\Omega$ is predominantly negative
when conditioned
on large values of $\Omega$, 
and thus consolidates 
the observed self-attenuation mechanism. 
Finally, it is worth noting that as $R/\eta$ becomes smaller
the curves for a given Reynolds number expectedly approach
the analytical limit given by Eq.~\ref{eq:slocal}.
The result for $R/\eta=0.5$ (not shown), 
was found to be virtually indistinguishable
from the corresponding curve showing the analytical
limit. 
}
%{In fact, we find that the average of the absolute value,
%$| P^L_\Omega |$ conditioned on $\Omega$, differs by no more than a few
%percents from $ - \langle  P^L_\Omega | \Omega \rangle$ when
%$\Omega /\langle \Omega \rangle$ is large, implying that the configurations
%where $P^L_\Omega$ are positive are negligible.}
%Although not shown, we also calculated the skewness 
%(normalized third moment), 
%%of $P^L_\Omega$, and  
%found that its behavior very closely follows
%the result in Fig.~\ref{fig:local}a.

The observation from Figs.\ref{fig:1} and \ref{fig:local}
that extreme vorticity fluctuations are accompanied by 
negative values of $P^L_\Omega$ indicates that the strain induced
locally acts to 
prevent further growth of enstrophy.
It is important to realize that this mechanism
is separate from viscous diffusion or dissipation of enstrophy \cite{BPB2020},
but still acts in conjunction with it.
Additionally, this self-attenuating mechanism is far stronger 
than a mere reduction (depletion) of 
non-linearity \cite{MoffTsi92,Gibbon:2014}.
\DB{
Depletion of non-linearity essentially refers
to weakening of vortex stretching 
(compared to its maximum possible amplitude)
\cite{tsi99}, 
which is evidently reflected in alignment of 
vorticity with intermediate eigenvector of
strain tensor and hence the weak curvature
of vortex tubes \cite{Ashurst87,Jimenez92} -- as also seen in 
Fig.~\ref{fig:1}a-b. 
}
However, the presence of self-attenuation 
suggests that the non-linearity itself 
could be capable of preventing a runaway blowup,
even as viscosity gets small
(as suggested by 
Fig.~\ref{fig:local}a, where the increase of $P_\Omega^L$
is merely shifted to larger values of $\Omega$ as
viscosity decreases).
A careful mathematical analysis of this mechanism 
and determining mathematical bounds on $P_\Omega^L$ could possibly
reveal a path in establishing global regularity of INSE.

\begin{figure}
\begin{center}
\includegraphics[width=0.45\textwidth]{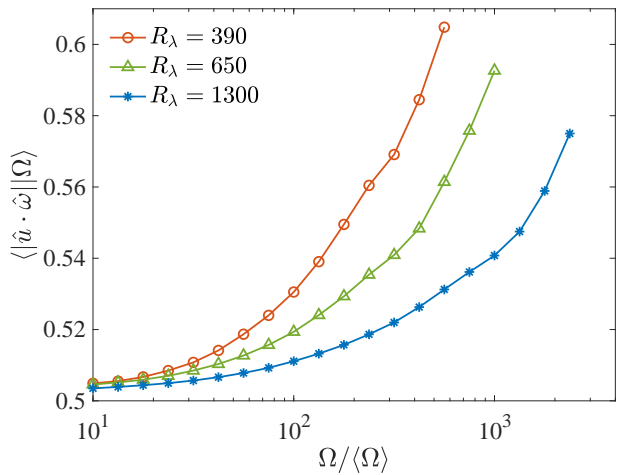}
\caption{
{\bf Preferential alignment of vorticity and velocity in regions of intense vorticity}.
Averaged absolute value of the cosine 
between velocity and vorticity vectors, conditioned on enstrophy 
relative to its mean value,
at Taylor-scale Reynolds numbers $\re=390-1300$.
}
\label{fig:heli}
\end{center}
\end{figure}

\paragraph*{\bf Connection to helicity:}
The presence of negative local stretching accompanying
intense vorticity raises additional questions
about the local flow structure.
Given that intense vorticity is arranged 
in tubes with weak curvature, additional
insight could be obtained by a simple
kinematic analysis of stretching generated by  such
structures. 
To this end, we consider a simple
axisymmetric vortex tube with a radius of curvature $R_c$ 
\cite{Siggia85,PumSig87,MoffKim19}.
Utilizing a curvilinear polar coordinate system: 
($\hat{\ve r}$, $\hat{\ve \theta}$, $\hat{\ve s}$), which respectively
correspond to unit vectors in the radial direction, the azimuthal 
direction and the direction tangent along the (curved) axis of the tube,
we assume that the vorticity is of the form 
$\ve \omega = \omega_s(r,s) \, \hat{\ve s} + \omega_\theta(r,s) \, \hat{\ve \theta}$.
The component $\omega_s$ corresponds to azimuthal velocity in the tube
similar to a two-dimensional Burgers vortex \cite{Burgers48},
whereas the component $\omega_\theta$ comes from
axial velocity along the tube.
Thereafter, 
utilizing Eq.~\eqref{eq:slocal},
one can derive (as shown in the Supplementary):
\begin{equation}
P^L_\Omega (R) = -\frac{R^2}{10} \left[ \mathcal{F} \{\omega_s, \omega_\theta \} + 
\mathcal{G} \{\omega_s, \omega_\theta\} 
\frac{\cos \theta}{R_c} \right] + \mathcal{O}(R^4) \ ,
\label{eq:Str_expr}
\end{equation}
which gives the local stretching induced by the vortex tube
as sum of two terms, involving 
$\mathcal{F}$ and $\mathcal{G}$, which are functions of 
$\omega_s$ and $\omega_\theta$ 
and their derivatives.

The term with a $\cos\theta/R_c$ dependence 
results from the curvature of the tube and produces
a dipolar structure, 
with positive and negative contributions 
depending on the sign of $\cos\theta$ \cite{pumir:1990} --
consistent with the structure seen in Fig.~\ref{fig:1}g and i.
In contrast, the term independent of $\cos\theta$ acts as a monopole.
Based on the results shown in Figs.~\ref{fig:1} and
\ref{fig:local}, the sign of $\mathcal{F}$ must be {\em positive}, 
and would result in 
attenuation of intense vorticity by the local strain.
Interestingly, $\mathcal{F}$ 
is identically zero if the component $\omega_\theta$
vanishes, i.e. there is no axial flow velocity.
This suggests that some local alignment between
vorticity and velocity must
occur when vorticity is large.
\DB{
Interestingly, a similar conclusion can also be reached
by realizing that the non-linear terms
in INSE, in Eq.~\eqref{eq:vort}, can be rewritten 
as $\nabla \times (\uu \times \ww)$.
Thus, local Beltramization, i.e., 
alignment of $\uu$ and $\ww$ 
in regions of large enstrophy would essentially
act to restrict the non-linear amplification 
\cite{MoffTsi92}. 
}

The above prediction 
is consistent with earlier results at low $\re$ 
\cite{Choi:09}, as well as with our own results at significantly
higher $\re$ in 
Fig.~\ref{fig:heli},
which shows the conditional average of the cosine between
velocity and vorticity, conditioned on enstrophy.
The average is taken over the absolute value, since the 
sign of the cosine is immaterial to measure the degree of
Beltramization (note that the dot product
of velocity and vorticity
is not sign-definite). 
For small values of $\Omega$, the average stays constant at $0.5$,
consistent with a uniform distribution of the cosine. 
However, the conditional average increases at large
$\Omega$, in good correlation 
with the increase of the magnitude of 
$P^L_\Omega$ seen in Fig.~\ref{fig:local}a.
Thus, in fully developed turbulence,
the intense whirling motions (vortex tubes),
emblematic of the small-scale structures, are
innately {\em three-dimensional} and {\em helical}.

\section*{Discussion}

\DB{
In conclusion,
we have utilized very well resolved numerical simulations
of fully developed turbulence to investigate
extreme fluctuations of vorticity,
which can be considered as signatures of potential
singularities of INSE.
Our results show that when vorticity is strongly amplified,
the non-linearity in its local neighborhood
remarkably counteracts further amplification, instead of enhancing it. 
In addition, this effect gets stronger as vorticity gets stronger
and also as Reynolds number increases (or viscosity decreases).
Thus, our results suggest that the non-linearity
-- which is responsible for amplification in the first place --
also encodes a mechanism 
(in conjunction with viscosity), which can
prevent a finite-time singularity from occurring.
A deeper understanding of this self-attenuation mechanism
based on a clear
physical argument could help to set stronger mathematical 
bounds on the stretching of vorticity \cite{Constantin:96,Deng:06},
and could be an essential ingredient to
prove global regularity of the INSE \cite{Fefferman}.  
}

\DB{
Another important observation in this regard is
the local Beltramization of the flow in 
regions of large enstrophy 
-- highlighting the helical nature of the small-scales 
of turbulence
(which are structurally arranged in vortex tubes).
While it was anticipated that reduction (depletion) of non-linearity
would lead to such helicity \cite{MoffTsi92}, 
the uncovered self-attenuation mechanism shows that the effect
is in fact much stronger, and directly counteracts 
vorticity amplification.
A promising direction in this regard could be
to extend the ideas based on helical decomposition 
to further analyze this local Beltramization
\cite{ConstMajda:88,waleffe93}.
Indeed, a recent work has established global regularity 
for a decimated
version of INSE which enforces helicity
to be sign-definite \cite{bt13}. 
A possible extension to full INSE,
in light of the uncovered self-attenuation mechanism,
presents an important challenge for future work.
}

\DB{
Finally, it is worth noting
that our numerical simulations of
stationary isotropic turbulence do not address
specific initial value problems,
such as those involving collisions between two
or more vortex tubes~\cite{pumir:1990,kerr:1993,luo14,Bren+16}.
Such special flow configurations are routinely 
studied to investigate the development of a possible
finite-time singularity, mostly 
in the context of inviscid
flows ($\nu=0$), i.e., the Euler equations. 
However, a conclusive demonstration of a blowup or lack thereof
still remains elusive \cite{gibbon08}.
While complicated interactions between vortex tubes
already occur in our simulations, it remains to 
be understood how the ideas developed here
would apply to these special configurations.
}

%
% --- Methods -----
%

\section*{Methods}

\paragraph*{\bf Direct numerical simulations:} 
The data utilized in the current work
are generated through direct numerical simulations (DNS)
of the incompressible Navier-Stokes equations (INSE) 
\begin{align}
\partial \uu/\partial t + \uu \cdot \nabla \uu = -\nabla P/\rho  + \nu \nabla^2 \uu + \ff \ ,
\end{align}
where $\uu$ is the divergence free velocity field ($\nabla \cdot \uu = 0$), 
$P$ is the pressure, $\rho$ is the fluid density, 
$\nu$ is the kinematic viscosity,
and $\ff$ corresponds to large scale forcing used to maintain a
statistically stationary state \cite{EP88}. 
The equations are solved 
using a massively parallelized version of the well-known
Fourier pseudo-spectral algorithm of Rogallo (1981) \cite{Rogallo}.
The aliasing errors resulting from the convolution sums
are controlled by grid shifting and spherical truncation \cite{PattOrs71}.
Our DNS corresponds to the canonical setup
of homogeneous and isotropic turbulence
with periodic boundary conditions on a cubic domain 
of side length $L_0=2\pi$,
which is ideal for 
studying small scales and hence extreme events at highest Reynolds numbers 
possible \cite{Ishihara09}.
The domain is discretized using $N^3$ grid points,
with uniform grid spacing $\Delta x = L_0/N$ in each direction.
We utilize explicit second-order Runge-Kutta for time integration,
where the time step $\Delta t$ is subject to the Courant number ($C$)
constraint for numerical stability: $\Delta t = C \Delta x/ ||\uu||_\infty$
(where $|| \cdot ||_\infty$ is the $L^\infty$ norm).

\begin{table}[h]
\centering
    \begin{tabular}{cccccc}
\hline
    $\re$   & $N^3$    & $k_{max}\eta$ & $T_E/\tau_K$ & $T_{sim}$ & $N_s$  \\
\hline
    140 & $1024^3$ & 5.82 & 16.0 & 6.5$T_E$ &  24 \\
    240 & $2048^3$ & 5.70 & 30.3 & 6.0$T_E$ &  24 \\
    390 & $4096^3$ & 5.81 & 48.4 & 4.0$T_E$ &  35 \\
    650 & $8192^3$ & 5.65 & 74.4 & 2.0$T_E$ &  40 \\
   1300 & $12288^3$ & 2.95 & 147.4 & 20$\tau_K$ &  18 \\
%    ~   & ~    & ~ & ~ & ~ & ~ & ~ \\
\hline
    \end{tabular}
\caption{Simulation parameters for the DNS runs
used in the current work: 
the Taylor-scale Reynolds number ($\re$),
the number of grid points ($N^3$),
spatial resolution ($k_{max}\eta$), 
ratio of large-eddy turnover time ($T_E$)
to Kolmogorov time scale ($\tau_K$),
length of simulation ($T_{sim}$) in statistically stationary state
and the number of instantaneous snapshots ($N_s$) 
used for each run to obtain the statistics.
}
\label{tab:param}
\end{table}

%As pointed out recently, studying small scales of turbulence
%requires stringent numerical resolution, 

The DNS database used in the current work
is summarized in Table~\ref{tab:param}, along with the main
simulation parameters.
An important consideration in studying
extreme events is that of spatial resolution,
which is measured in pseudo-spectral DNS by the parameter
$k_{max}\eta$, where $k_{max}=\sqrt{2}N/3$ is the maximum resolved
wavenumber on a $N^3$ grid and $\eta$ is the Kolmogorov length scale.
Equivalently, one can use the ratio $\Delta x/\eta$ which is 
approximately equal to $3/k_{max}\eta$.
The runs with Taylor-scale Reynolds numbers, $\re$,
in the range $140 \le \re \le 650$ were also utilized in our
recent work \cite{BPBY2019} and all have a very high spatial
resolution,
$k_{max}\eta\approx6$ (or $\Delta x/\eta \approx 0.5$).
This resolution should be compared
to the one used in comparable numerical investigations of turbulence
at high Reynolds numbers, which are mostly in the range
$1 \le k_{max} \eta \le 1.5$ \cite{Ishihara09,BSY.2015} -- which do not
resolve the extreme events adequately.
%As established in our recent work \cite{BPBY2019},
%such a resolution is more than adequate
%to accurately resolve and study the small-scales at
%the $\re$ considered, although the resolution constraints
%appear to be more severe as the Reynolds number increases.
In addition to our previous runs, 
we have performed a new run at
significantly higher $\re$ of $1300$,
on a larger $12288^3$ grid with a small-scale 
resolution of $k_{max}\eta=3$ 
(or $\Delta x/\eta\approx1$).
This is one of the largest DNS reported
to date -- comparable with \cite{Ishihara16} which also reported
results from $12288^3$ run at $\re=2300$, but with 
$k_{max}\eta\approx1$ (where the small-scales were not properly
resolved).

We have also listed the simulation length $T_{sim}$ used for generating 
independent ensembles, 
in terms of the large-eddy turnover time ($T_E$) or the 
Kolmogorov time scale ($\tau_K$). 
The statistical results are obtained by averaging over $N_s$
independent ensembles, which are uniformly spread out over the 
simulation length. 
Note, the range of time scales
is typically given by the ratio $T_E/\tau_K$, which scales linearly
with $\re$ \cite{tl72}. However, the time scale of extreme events
which we consider here is smaller than $\tau_K$,
getting even smaller as $\re$ increases \cite{BPBY2019}.

%
% --- other stuff -----
%

\section*{Data availability}
The data that support the findings of this study are 
available from the corresponding author on request.

\section*{Code availability}
The simulation and post-processing codes that have been 
used to produce the results of this study are available 
from the corresponding author on request.

\section*{Acknowledgements}

We gratefully acknowledge the Gauss Centre for Supercomputing e.V.
(www.gauss-centre.eu) for providing computing time on the
GCS supercomputers JUQUEEN and JUWELS at J\"ulich Supercomputing Centre (JSC),
where the simulations reported in this paper were performed.
We acknowledge support from the Max Planck Society.
We also thank P. K. Yeung for sustained collaboration
and partial support 
under the Blue Waters computing project
at the University of Illinois Urbana-Champaign.

\section*{Author contributions}

D.B. performed the numerical simulations and data analyses. 
All authors designed the research and interpreted the data.
D.B. and A.P. wrote the manuscript, and E.B. commented on it.

\section*{Additional information}

The authors declare no competing financial interests.
The manuscript is accompanied by a Supplementary.

%\bibliographystyle{apsrev4-1}
%\bibliography{large_grad}


\end{document}